\documentclass[aps,prb,twocolumn,showpacs]{revtex4}%
\usepackage{amsfonts}
\usepackage{amsmath}
\usepackage{amssymb}
\usepackage{graphicx}%
\begin{document}
\title{I-V curves and intergranular flux creep activation energy in the magnetic
superconductor RuSr$_{2}$GdCu$_{2}$O$_{8}$}
\author{S. Garc\'{\i}a}
\affiliation{Laboratorio de Superconductividad, Facultad de F\'{\i}sica-IMRE, Universidad
de La Habana, San L\'{a}zaro y L, Ciudad de La Habana 10400, Cuba.}
\author{L. Ghivelder}
\affiliation{Instituto de F\'{\i}sica, Universidade Federal do Rio de Janeiro, C.P. 68528,
Rio de Janeiro , RJ 21941-972, Brazil}
\pacs{74.72.-h, 74.25.Fy, 74.50.+r}

\begin{abstract}
A systematic study of I-V characteristic curves for RuSr$_{2}$GdCu$_{2}$%
O$_{8}$ [Ru-(1212)] is presented, with magnetic fields up to 3 T and 5 K
$\leq$ $T$ $\leq$ 30 K, in the region of the superconducting transition. The
activation energy $E_{a}(H,T)$ for flux line depinning was determined by
fitting the nonlinear region of the curves using the flux creep model.
$E_{a}(H,T)$ was found to vary linearly with temperature, while a power-law
dependence on the magnetic field was observed up to $H$ = 0.1 T, where an
abrupt reduction in its decreasing rate occurs. The extrapolated value,
$E_{a}(0,0)$ $\simeq$50 meV, is twice the reported value for YBa$_{2}$Cu$_{3}%
$O$_{7-\delta}$, but the critical current density $J_{C}(0,0)$ $\simeq$70
A/cm$^{2}$ is about one order of magnitude lower. These results are explained
as a consequence of the contribution of the magnetization in the grains to the
effective field at the intergranular links and to a spin-flop transition of
the Ru-sub-lattice.

\end{abstract}
\received[Received text]{1 November 2003}

\maketitle

The ruthenate-cuprates RuSr$_{2}$RCu$_{2}$O$_{8}$ , where R = Gd or Eu, are
superconducting (SC) materials which exhibits long-range ferromagnetic (FM)
order at a temperature $T_{M}$ $\approx$ 135 K, well above the onset
temperature of the SC transition, at $T_{SC}$ $\approx$ 40 K.\cite{Tallon}
Although since the early works there were evidences of interaction between the
Ru moments and the charge carriers, the interplay between magnetism and the
transport properties in this compound is still an open issue. Different
magneto-resistive behaviors have been reported above and below $T_{M}%
$.\cite{McCrone} Peaks in the dc and microwave resistivity\cite{Pozek01} and
changes in the Hall coefficient near $T_{M}$ have been also
observed.\cite{Pozek02} In relation to the SC transition, it has been found
that the resistive curves show a quite different magnetic field dependence
from that observed in the conventional cuprates.\cite{Chen} Another important
characteristic is the large SC transition width ($\Delta T_{C}$ $\sim$ 15-20
K) even in good quality polycrystalline compounds. Microwave resistance
measurements in powder samples have demonstrated that this is due to a
significant contribution from the intergrain network, which is assumed to be
under the influence of spontaneous vortex phase formation within the
grains.\cite{Pozek01} Thus, a study of the magnetic field dependence of the
intergrain coupling is essential to determine if the magnetization in the
grains actually leaves a sizeable effect in the connectivity of the weak-link
network. However, a quantitative study of the magnitudes characterizing the
intergrain transport properties in the ruthenate-cuprates is still lacking.

This motivated our investigation of the $I$-$V$ dissipation curves in
RuSr$_{2}$GdCu$_{2}$O$_{8}$ [Ru-(1212)]. In this work we present a systematic
study of the magnetic field and temperature dependence of the characteristic
$I$-$V$ curves in this compound, measured in fields up to $H$ = 3 T. The
intergranular flux creep activation energy of the Josephson vortices $E_{a}$
and its dependence on the applied current, magnetic field and temperature were
determined by fitting the non linear region of $I$-$V$ curves according to the
flux creep model.\cite{Yeshurun} A comparison with the conventional cuprates
concerning the contributions to $E_{a}(H,T)$ and the behavior of the
intergrain transition peak, as determined from the resistive curves, suggest
that the Ru magnetization makes an important contribution to the local field
at the junctions. This is strongly supported by the $E_{a}(H)$ behavior,
showing an abrupt reduction, by a factor of $\sim$7 times, in the decreasing
rate at $H$ = 0.1 T. We argue that this is a consequence of a spin-flop
transition of the Ru moments.

Polycrystalline RuSr$_{2}$GdCu$_{2}$O$_{8}$ was prepared by conventional
solid-state reaction; details of sample preparation and microstructure can be
found elsewhere.\cite{Garcia} The room temperature x-ray diffraction pattern
corresponds to RuSr$_{2}$GdCu$_{2}$O$_{8}$, with no spurious lines being
observed. Scanning electron microscopy revealed a relatively uniform size
distribution with rounded grains of about 1-3 $\mu$m.\cite{Garcia} Bars of
$\sim$10 mm in length and 0.6 mm$^{2}$ cross sectional area were cut from the
sinterized pellet. Resistive measurements were made with a standard four-probe
technique, using silver paint with a contact resistance of $\sim$1-2 $\Omega$.
Current vs. voltage ($I$-$V$) measurements were performed at various
temperatures and magnetic fields in a Quantum Design PPMS system, with $H$ =
0, 0.01, 0.03, 0.05, 0.1, 0.3, 1, and 3 T, for 5 K $\leq$ $T$ $\leq$ 30 K.
Since the zero resistance temperature $T_{R=0}$ rapidly diminishes with the
increase in $H$, the accessible temperature measuring range below T$_{R=0}$
was about 10-15 K for $H$ $<$ 0.3 T and $\sim$\ 5-8 K for higher fields. For
each magnetic field, a set of curves was measured with a temperature interval
$\Delta$T = 0.3 K and a current step of 0.45 mA. The usual electric field
criterion of 1 $\mu$V/cm was used to determine the critical current density
$J_{C}$.\cite{Peterson, Altshuler} Typically, about 30-40 points were
collected in the nonlinear region.

Figure 1 shows a set of experimental $I$-$V$ curves measured for $H$ = 0.3 T.
A similar behavior was observed for other measured magnetic fields. The
continuous lines are fittings described below. These curves exhibit
qualitatively the same features observed for the conventional granular
cuprates: a noticeable decrease of $J_{C}$ with the rise in temperature and a
clearly distinguishable nonlinear region which smoothly evolves to a linear
dependence. For $H$ = 0, the critical current density at the lowest measured
temperature ($T$ = 10 K) was only 50 A/cm$^{2}$. A linear extrapolation to $T$
= 0 K gives $J_{C}(T=0,H=0)$ = 70 A/cm$^{2}$. The non linear region of the
experimental curves was fitted using the well known expression,\cite{Yeshurun}%

\begin{equation}
V(I,H,T)=aI\exp[E_{a}(I,H,T)/kT] \tag{1}%
\end{equation}
where $I$ is the applied current, $k$ is the Boltzmann constant, and $E_{a}$
is the flux creep activation energy of the Josephson vortices. The pre-factor
$(aI)$ allows a continuous transition to the flux flow regime. The activation
energy was assumed to vary linearly with $I$ as $E_{a}(I,H,T)=E_{a}%
(H,T)(1-I/I_{0})$. An alternative dependence, $E_{a}$ $\varpropto$
$ln(1-I/I_{0})$, was also considered, leading to unsatisfactory fittings. For
a fixed magnetic field, the $E_{a}(H,T)$ values for a set of curves were
determined and plotted as a function of temperature, as shown in Fig.2. A
linear behavior was obtained for $H$ $\leq$ 0.3 T , where the measuring
temperature intervals are larger, with slopes which regularly diminish with
the increase in field. For $H$
$>$
0.3 T, although the studied temperature range is smaller, the data also
clearly follows a linear dependence. This indicates that the intergrain links
in Ru-(1212) behave as SIS-junctions. It is interesting to note that the
linear behavior is observed for temperatures relatively far from $T_{SC}$ (0.1
$\leq$ $T/T_{SC}$ $\leq$ 0.6). Due to the broad SC transition width, it is not
possible to study the behavior near $T_{SC}$. The extrapolation to $T$ = 0 of
the zero field curve gives $E_{a}(0,0)$ = 50 meV. The activation energies were
found to strongly depend on the intensity of the field. Figure 3 shows a
log-log plot of the field dependence of $E_{a}$, measured at $T$ = 10 K. It is
found that a power-law dependence, $E_{a}(H)$ $\varpropto$ $H^{-n}$, with $n$
= 0.75, describes quite well the data up to $H$ = 0.1 T, where a noticeable
reduction in the decreasing rate is observed.

A comparison of the obtained flux creep activation energies with the
conventional cuprates must be performed carefully. One important point is to
distinguish between the pinning potential $U$ and the activation energy for
flux line depinning\cite{Muller90}%
\begin{equation}
E_{a}(H,T)=U(H,T)-JBV_{b}d\tag{2}%
\end{equation}
where $J$ is the applied current density, $B$ is the local macroscopic flux
density, $V_{b}$ is the intergranular flux bundle volume, and $d$ is the
intergranular potential range. There has been a plethora of $U$ and $E_{a}$
energy values, ranging from a few eV's to a few meV's, obtained by a number of
different methods. At times, there has been a tendency to interchange freely
these parameters and to bunch all the values together, regardless of whether
one is dealing with single crystals or polycrystalline materials. So, we make
a comparison solely with reports which make this clear distinction. Since
Ru-(1212) is simply obtained from the YBa$_{2}$Cu$_{3}$O$_{7-\delta}$ (YBCO)
structure by full replacement of Cu(1) sites at the chains by Ru ions, which
add two oxygen atoms to their neighborhoods,\cite{McLaughlin} this material is
an ideal reference for comparison with the conventional cuprates. While the
$U(0,0)$ values in good quality YBCO ceramic samples are in the 10-15 eV
range,\cite{Nikolo94,Nikolo89,Muller90} $E_{a}(0,0)$ is only about 26 meV,
with $J_{C}(0,0)$ $\approx$ 600 A/cm2.\cite{Nikolo94} We note that although
our value of $E_{a}(0,0)$ = 50 meV is twice the reported value for YBCO,
$J_{C}(0,0)$ is about one order of magnitude lower. Although an independent
determination of the pinning potential $U$ is needed for an exact quantitative
balance of the different contributions to $E_{a}$, it is interesting to
analyze the second term in Eq. (2), since it depends on the local field at the
intergrain junctions. Magnetic fields of $\sim$600-700 Oe have been measured
by muon spin rotation\cite{Bernhard} and Gd-electron paramagnetic
resonance\cite{Fainstein} in sites located in the neighborhood of the
RuO$_{2}$ layers. Although the dipolar field rapidly decays with distance,
fields as small as a few tens of Oersted or less at the junctions, acting in
the whole volume of the sample, will have a great impact in the connectivity
of the weak-link network. Eventually, the effect of the local field may
completely inactivate low quality junctions, in such a way that the current
loops of the Josephson vortices can meander only through a fraction of high
quality connections. This leads to an increase in the area and volume of the
intergranular flux bundle. For polycrystalline samples, $V_{b}=\mathit{l}%
A_{b}$, where $\mathit{l}$ is the correlation length along an intergranular
flux line and $A_{b}$ is the area of the flux bundle. Since $J_{C}$ $\sim
$\ ($V_{b})^{-1}$,\cite{Nikolo94} a well-connected granular material with high
$J_{C}$ should be close to the minimal values $A_{b}\approx\pi(R_{g})^{2}$ and
$\mathit{l}=2R_{g}$, where $R_{g}$ is the average grain radius. This
assumption was proved to describe well the intergranular contribution to the
ac magnetic susceptibility in good quality YBCO\cite{Muller90} and
BSCCO\cite{Muller91} ceramic samples. In a network with its intergranular
coupling weakened by an internal field, as we believe to occur in Ru-1212, a
moderate increase in both $\mathit{l}$ and in the radius of the intergranular
current loop of only twice the minimal values can account for a decrease in
$J_{C}$ of about one order of magnitude, which is the variation observed in
comparison to YBCO. The assumption of an increased $V_{b}$ is supported by the
fact that the intergrain transition peak for Ru-1212, as determined from the
derivative of the resistive curve, is very well defined and intense, as shown
in the inset of Fig. 3, indicating that the intergrain transition occurs in a
narrow temperature interval across junctions of similar qualities, which
\textquotedblleft survive\textquotedblright\ the action of the magnetization
of the grains. This feature is at variance with the conventional cuprates,
exhibiting an intergrain peak of small amplitude as a consequence of a
broad-in-temperature phase-lock process across a wide distribution of links
qualities. The idea that the magnetization of the grains plays a relevant role
in the connectivity of the intergrain network is confirmed by the $E_{a}(H)$
dependence. We have previously found that in Ru-(1212) the ac magnetic
susceptibility in the presence of a dc bias and the intergrain transition
temperature change their behavior just at $H$ $\cong$ 0.1-0.3 T.\cite{Garcia}
These results were explained in terms of a re-arrangement of the Ru moments,
leading to a state of reduced magnetization. We recall here that a change in
the neutron diffraction pattern of Ru-(1212) is observed when a magnetic field
$H$ = 0.4 T is applied,\cite{Lynn} a result interpreted as due to a spin-flop
transition. Also, detailed magnetic measurements in this compound indicate
that a spin-flop transition should occur at a critical field of $\sim$0.14
T.\cite{Butera} These values are near to the field at which the change in the
$E_{a}(H)$ behavior is observed. We believe that the reduction in the
suppression rate of $E_{a}$ for $H$
$>$
0.1 T is a consequence of a magnetic transition of the Ru-sub-lattice, leading
to a decrease of the average contribution from the magnetization in the grains
to the net local field at the junctions.

In summary, we have shown that the intergranular dissipation curves for
RuSr$_{2}$GdCu$_{2}$O$_{8}$ are properly described by the flux creep model. A
detailed study of $I$-$V$ curves in a polycrystalline samples allowed, for the
first time to our knowledge, a quantitative determination of the flux creep
intergranular activation energy and its temperature and magnetic field
dependence in this compound. Evidence was provided indicating that the effect
of the magnetization of the grains on the weak-link network is essential to
understand the differences with the conventional high temperature
superconducting cuprates.

\begin{acknowledgments}
S. G. acknowledges financial support from FAPERJ. We thank J. Musa for sample preparation.
\end{acknowledgments}

\section{Figure Captions}

Figure. 1. Experimental $I$-$V$ curves measured for RuSr$_{2}$GdCu$_{2}$%
O$_{8}$ [Ru-(1212)]\ with $H$ = 0.3 T and 7.7 $\leq$ $T$ $\leq$ 11.9 K. The
temperature step between the selected curves in the graph is 0.6 K. The
continuous lines are fittings according to the flux creep model.

Fig. 2. Temperature dependence of the flux creep activation energy for
different applied magnetic fields. The magnetic fields are, from right to
left, $H$ = 0, 0.01, 0.03, 0.05, 0.1, 0.3, and 3 T. The curve for $H$ = 1 T
was omitted for clarity. The continuous lines are linear fittings.

Fig. 3. A log-log plot of the magnetic field dependence of the flux creep
activation energy at $T$ = 10 K. The continuous line is a guide to the eyes.
The dotted line is an extrapolation of the behavior for $H$ $\leq$ 0.1 T,
emphasizing the change in the decreasing rate for higher fields. Inset: intra-
($T_{1}$) and intergrain ($T_{2}$) transitions for Ru-(1212), observed in the
derivative of the resistive curve for $H$ = 0.

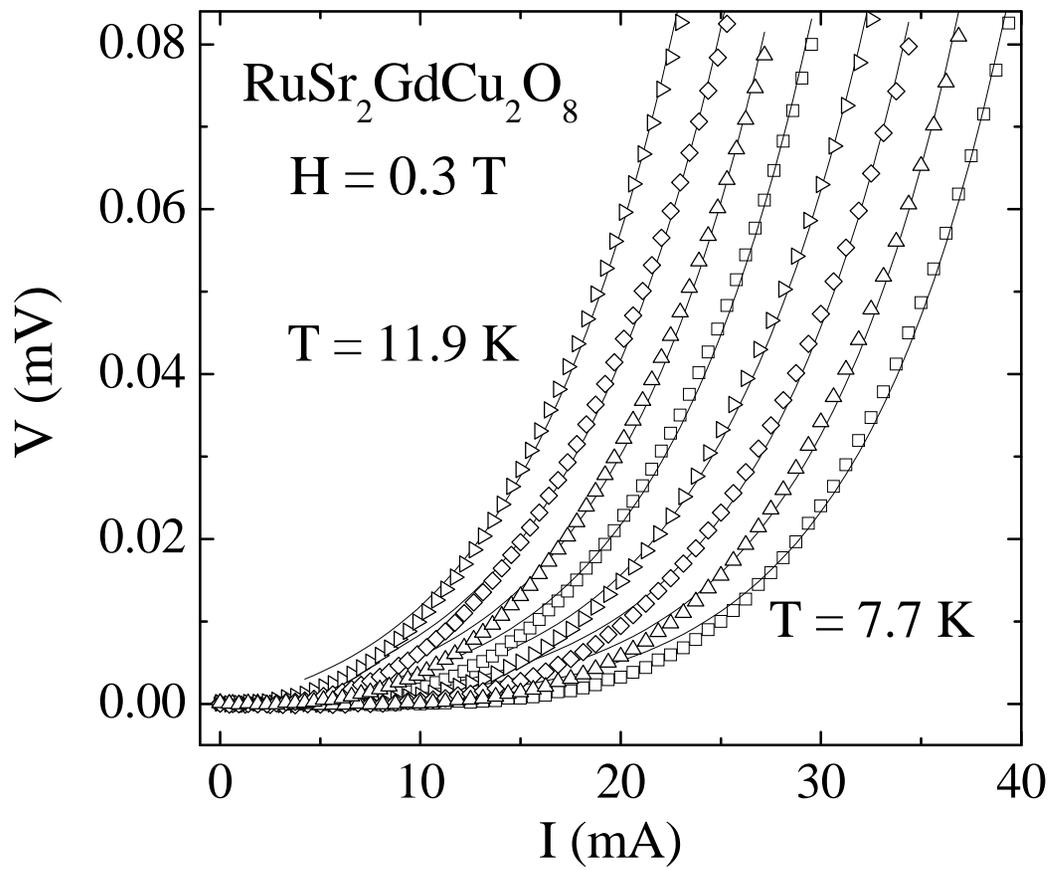
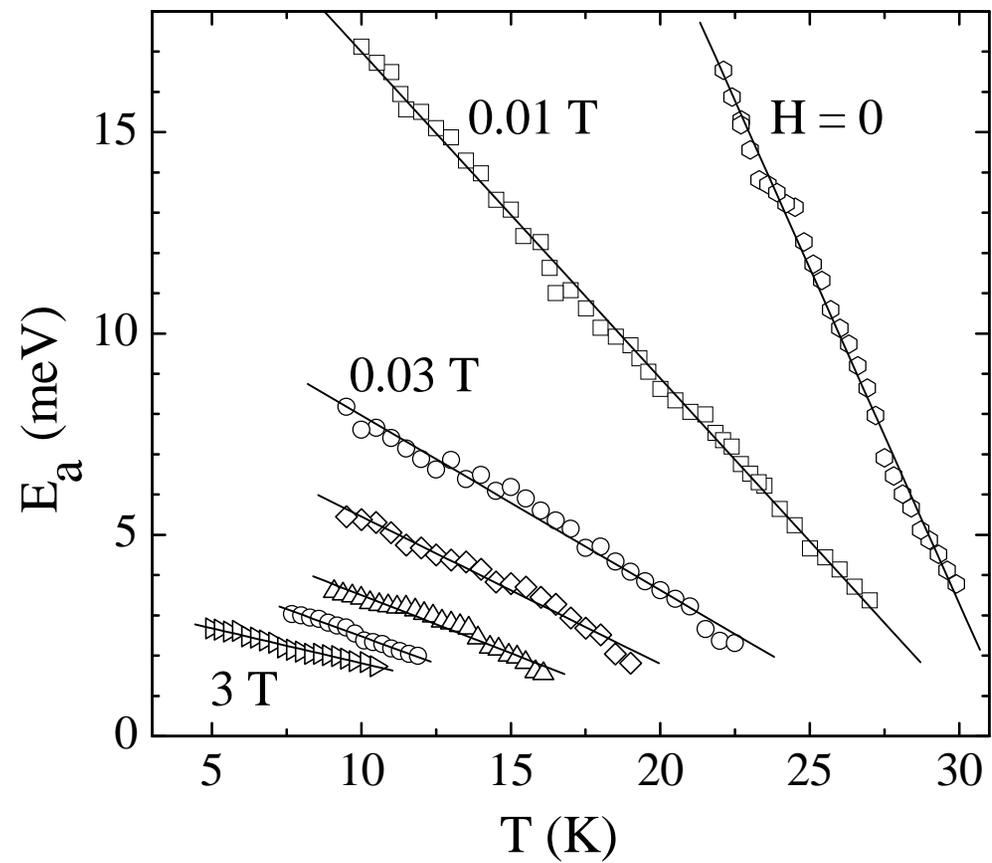
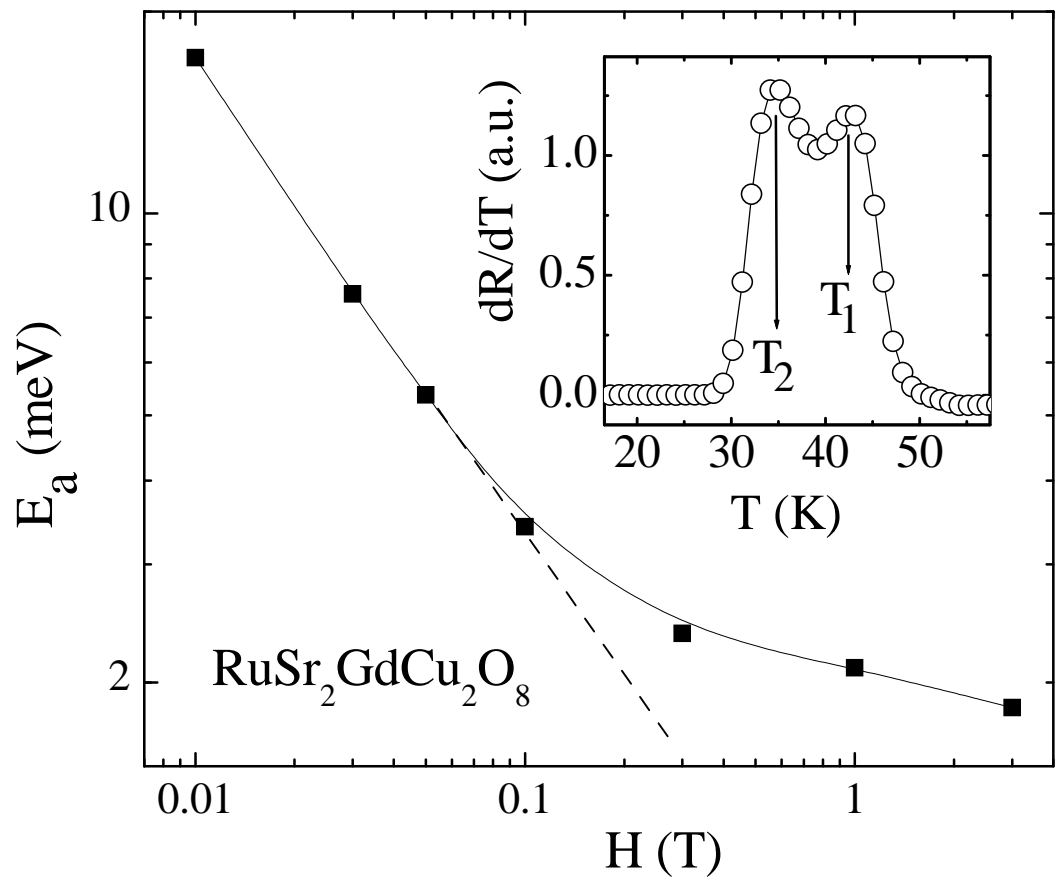
\end{document}